\documentclass[preprint]{aastex}
\usepackage{epsfig,amssymb}

\shorttitle{Gravitational Waves as electron accelerators}
\shortauthors{Vlahos et al.}

\begin{document}


\title{Impulsive electron acceleration by Gravitational Waves }


\author{Loukas Vlahos, George Voyatzis and
  Demetrios Papadopoulos}
\affil{Department of Physics,\\ University of Thessaloniki,
Thessaloniki 54421, Greece}


\begin{abstract}
We investigate the non-linear interaction of a strong
Gravitational Wave with the plasma during the collapse of a
massive magnetized star to form a black hole, or during the
merging of neutron star binaries (central engine). We found that
under certain conditions this coupling may result in an efficient
energy space diffusion of particles. We suggest that the
atmosphere created around the central engine is filled with 3-D
magnetic neutral sheets (magnetic nulls). We demonstrate that the
passage of strong pulses of  Gravitational Waves through the
magnetic neutral sheets accelerates electrons to very high
energies. Superposition of many such short lived accelerators,
embedded inside a turbulent plasma, may be the source for the
observed impulsive short lived bursts. We conclude that in several
astrophysical events, gravitational pulses may accelerate the tail
of the ambient plasma to very high energies and become the driver
for many types of astrophysical bursts.
\end{abstract}
\keywords{gravitational waves--compact objects:--- particle
acceleration.}

\section{Introduction}

The interaction of Gravitational Waves (GW) with the plasma and/or
the electromagnetic waves propagating inside the plasma, has been
studied extensively
\citep{DeW,Coop,Zel,Gerl,Gri,Den,MacThor,Dem,Dan,Bod,Mar00,Bro00,Ser1,Ser2,
Bro01,Moo} . All well known approaches for the study of the
wave-plasma interaction have been used, namely the Vlasov-Maxwell
equations \citep{Macedo}, the MHD equations \citep{PapEsp,Pap,Moo}
and the non-linear evolution of charged particles interacting with
a monochromatic GW \citep{Var}. The Vlasov-Maxwell equations and
the MHD equations were mainly used to investigate the linear
coupling of the GW with the normal modes of the ambient plasma,
but the normal mode analysis is a valid approximation only when
the GW is relatively weak and the orbits of the charged particles
are assumed to remain close to the undisturbed ones. Several
studies have also explored, using the weak turbulence theory, the
non-linear wave-wave interaction of plasma waves with the GW (see
\citet{Bro00}).

The strong nonlinear coupling of isolated charged particles with a
coherent GW was studied using the Hamiltonian formalism
\citep{Var,Klei1,Klei2}. The main conclusion of these studies was
that the coupling between GW and an isolated charged particle
gyrating inside a constant magnetic field can be very strong only
if the GW is very intense. This type of analysis can treat the
full non-linear coupling of the charged particle with the GW but
looses all the collective phenomena associated with the excitation
of waves inside the plasma and the back reaction of the plasma
onto the GW.

In this article, we re-investigate the non-linear interaction of
an electron with a GW inside a magnetic field, using the
Hamiltonian formalism.  Our study is applicable at the
neighborhood of the central engine  (collapsing massive magnetic
star \citep{Fry,Dim,Baum}) or during the final stages of the
merging of neutron star binaries \citep{Ruf,Shi}. We find that a
strong but low frequency ($10$ KHz) GW can resonate with ambient
electrons only in the neighborhood of magnetic neutral sheets and
accelerates them to very high energies in milliseconds.
Relativistic electrons travel along the magnetic field, escaping
from the neutral sheet to the super strong magnetic field, and
emitting synchrotron radiation. We propose that the passage of a
GW through numerous localized neutral sheets will create spiky
sources which collectively produce the highly variable in time.

In section 2, we analyze the interaction of  a GW with a single
electron inside a constant magnetic field. In section 3 we
investigate the diffusion to high energies of a distribution of
electrons inside the GW. In section 4, we propose a new model for
strong coupling of the GW with the turbulent plasma at the
atmosphere of the central engine and the resulting impulsive
synchrotron emission and finally, in section 5, we summarize our
results.

\section{The Hamiltonian formulation of the GW-particle interaction}
The motion of a charged particle in a curved space and in the
presence of a magnetic field is described by a Hamiltonian, which,
in a system of units $m=c=G=1,$ is given by

\begin{equation}
H(x^{\alpha},p_{\alpha})=\frac{1}{2}g^{\mu\nu}(p_\mu-eA_\mu)(p_\nu-eA_\nu)
=\frac{1}{2},\;\; \alpha,\mu,\nu=0,...,3. \label{Hoo}
\end{equation}
$g^{\mu\nu}=g^{\mu\nu}(x^\alpha)$ are the contravariant components
of the metric tensor of the curved space and
$A_\mu=A_\mu(x^\alpha)$ are the components of the vector potential
of the magnetic field \citep{Misner}. The variables $p_\alpha$ are
the generalized momenta corresponding to the coordinates
$x^\alpha,$ and their evolution with respect to the proper time
$\tau$ is given by the canonical equations
\begin{equation}
\frac{d x^{\alpha}}{d \tau}=\frac{\partial H}{\partial
p_{\alpha}}, \:\:\:\: \frac{d p_{\alpha}}{d \tau}=-\frac{\partial
H}{\partial x^{\alpha}}. \label{HamEqs}
\end{equation}

We assume a constant magnetic field $\vec{B}=B_0 \vec{e}_z$ which
is produced by the vector potential
\begin{equation}
A_0=A_1=A_3=0,\;A_2=B_0 (x^1+c_0),\;c_0:\textrm{const.,}
\label{MField}
\end{equation}
and that a GW propagates in a direction $\vec{k}$ of angle
$\theta$ with respect to the direction of the magnetic field. In
that case the nonzero components of the metric tensor are (see
\citet{Ohanian,PapEsp}) $g^{00}=1$ and
\begin{equation}
\begin{array}{ll}
g^{11}=\frac{1-a \sin^2\theta \cos\psi}{-1+a \cos\psi} &
g^{22}=\frac{-1}{1+a\cos\psi} \\ g^{33}=\frac{1-a \cos^2\theta
\cos\psi}{-1+a \cos\psi} & g^{13}=g^{31}=\frac{(-a/2) \sin 2\theta
\cos\psi}{-1+a \cos\psi},
\end{array} \label{MTensor}
\end{equation}
where $a$ is the amplitude of the GW and $\psi=k_\mu x^\mu=\nu
(\sin\theta x^1+\cos\theta x^3 - x^0).$ The parameter $\nu$ is the
relative frequency of the GW, i.e. $\nu=\omega/\Omega$, where
$\Omega=e B_0/mc$ is the Larmor angular frequency. We make use of
the scaling $eB_0=1$, thus $\Omega=1$.

In the above formalism, the coordinate $x^2$ is ignorable, so
$p_2=const.$. By setting the constant $c_0$ in Eq. (\ref{MField})
equal to $p_2$ we get an appropriate gauge that reduces by one
degree of freedom the Hamiltonian (Eq. (\ref{Hoo})), which takes
the form
\begin{equation}
H=\frac{1}{2}\left ( p_0^2-\frac{1-a s_\theta^2
\cos\psi}{1-a\cos\psi}p_1^2-\frac{1-a c_\theta^2
\cos\psi}{1-a\cos\psi}p_3^2+\frac{2 \alpha s_\theta c_\theta
\cos\psi}{1-a\cos\psi}p_1 p_3 - \frac{x_1^2}{1+a\cos\psi} \right
),
 \label{Ho1}
\end{equation}
where we use the notation $c_\theta=\cos\theta$ and
$s_\theta=\sin\theta$ for brevity. We apply the canonical
transformation of variables $(x^0,x^1,x^3,p_0,p_1,p_3) \rightarrow
(\chi,q,\phi,I,p,J)$ using the generating function
\begin{equation}
F(x^0,x^1,x^3,I,p,J)=x^0 I+x^1 p + \nu(s_\theta x^1+c_\theta
x^3-x^0)J. \label{GenF}
\end{equation}
The relation between the old and the new variables is given by the
equations
\begin{equation}
\begin{array}{ll}
\chi=x^0, & I=p_0+p_3/c_\theta \\
q=x^1, & p=p_1-(s_\theta/c_\theta)p_3\\
\phi=\nu(s_\theta x^1+c_\theta x^3-x^0) & J=p_3/(c_\theta
s_\theta). \end{array} \label{Vtrans}
\end{equation}
In the new variables the Hamiltonian (Eq. (\ref{Ho1})) takes the
form
\begin{equation}
H=\frac{1}{2}\left ( I^2-2I \nu J -2 s_\theta \nu J p - \frac{1-a
s_\theta^2 \cos\phi}{1-a\cos\phi} p^2 - \frac{q^2}{1+a \cos\phi}
\right). \label{Ho2}
\end{equation}
Since the variable $\chi$ is ignorable, $I$ is a constant of
motion and Eq. (\ref{Ho2}) can be studied as a system of two
degrees of freedom, where $I$ is a parameter. The variables $q$
and $p$ are associated with the gyro-motion. $H$ is of mod($2\pi$)
with respect to the angle-variable $\phi$ and the variable $J$ is
related linearly with the energy $\gamma=(1-\upsilon^2)^{-1/2}$ of
the particles according to the equation
\begin{equation}
\gamma=I-\nu J \label{gammaJ}.
\end{equation}
The equations of motion are
\begin{equation}
\begin{array}{ll}
\dot{q}=-s_\theta \nu J- \frac{1-a s_\theta^2
\cos\phi}{1-a\cos\phi} p &\hspace{0.5cm} \dot{p}=\frac{q}{1+a \cos\phi} \\
\dot{\phi}=-\nu I-s_\theta \nu p & \hspace{0.5cm}
\dot{J}=\frac{a}{2}\left ( \frac{q^2}{(1+a\cos\phi)^2} -
\frac{c_\theta^2 p^2}{(1-a \cos\phi)^2} \right ) \sin\phi,
\end{array} \label{EqM}
\end{equation}
where the dot means derivative with respect to the proper time
$\tau$.  Furthermore, Eq.(\ref{Ho2}) can be written as a perturbed
Hamiltonian in the usual way, i.e.
\begin{equation}
H=H_0+a H_1 + a^2 H_2 + ..., \label{Ho3}
\end{equation}
where
\begin{equation}
H_m=-(c_\theta^2 p^2+(-1)^m q^2) \cos^m\phi,\:\:m\geq 1
\label{Hpert}
\end{equation}
are the perturbation terms and
\begin{equation}
H_0=\frac{1}{2}(I^2-2I\nu J-2 s_\theta\nu Jp)-\frac{1}{2}(p^2+q^2)
\label{H0}
\end{equation}
is the integrable part of the system that describes the
unperturbed helical motion of the particle in the flat space.
Considering action-angle variables $(J_1,J_2,\phi_1,\phi_2)$,
Eq.(\ref{H0}) takes the form
\begin{equation}
H_0(J_1,J_2)=\frac{I}{2}-I \nu J_1+\frac{s_\theta^2 \nu^2}{2}
J_1^2-J_2, \label{H0J}
\end{equation}
where $\phi_1=\phi, J_1=J$ and
\[
J_2=\frac{1}{2\pi} \oint pdq = \frac{1}{2} (I^2-2I\nu J
+s_\theta^2 \nu^2 J^2-1)\:,\:\:\phi_2=\arcsin\left ( \frac{\pm
q}{\sqrt{2 J_2}} \right ).
\]
Therefore, the unperturbed system is isoenergeticaly
non-degenerate for $\theta\neq 0$ \citep{ArnoldIII} and the gyro
motion of the particles is represented by trajectories that twist
invariant tori with angular frequencies $\omega_1=\partial
H_0/\partial J_1$ and $\omega_2=\partial H_0/\partial J_2$. The
periodic or quasi-periodic evolution of the trajectories depends
on whether the rotation number, defined by
\begin{equation}
\rho=\frac{\omega_1}{\omega_2}=\nu I-s_\theta^2 \nu^2
J_1=\nu(c_\theta^2 I +s_\theta^2 \gamma), \label{RotN}
\end{equation}
is rational or irrational, respectively.

Most of the  invariant tori will persist with the presence of the
perturbation introduced by the GW, if the amplitude is suficiently
small, according to the KAM theorem \citep{ArnoldIII}. The orbits
of the particles remain close to the unperturbed ones but their
projection on the $x^1-x^2$ plane is not exactly circular and
periodic. Close to the resonant tori, where $\rho$ is rational,
the Poincar\'e-Birkhoff theorem applies; a finite number of pairs
of stable and unstable periodic trajectories survive, producing
locally a pendulum like topology in phase space \citep{SUZ}.

\begin{figure}
\centering
\includegraphics[width=14cm]{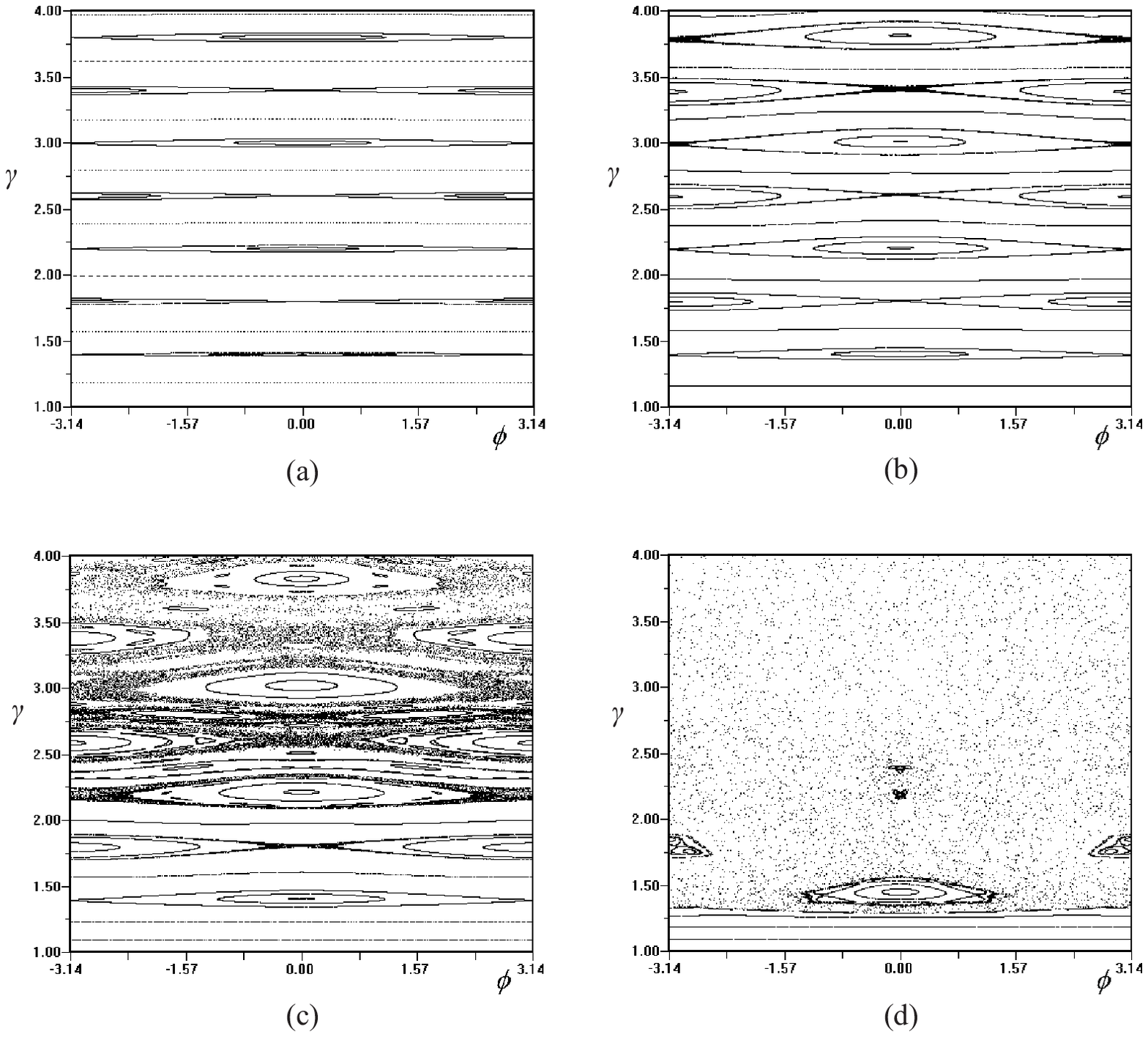}
\caption{Typical Poincar\'e sections on the plane ($\phi,\gamma$)
of the perturbed system for $\nu=5, \theta=45^o$ and a) $a=0.001$
b) $a=0.01$ c) $a=0.02$ and d) $a=0.1$.} \label{PNC1}
\end{figure}

Since the system is of two degrees of freedom, we can study its
evolution by using the Poincar\'e sections
$P_S=\{(\phi,\gamma),q=0,H=1/2\}$ choosing specific sets of the
parameters $a,I,\nu$ and $\theta$. In the numerical calculations,
which will follow, we set $I=1$. For the unperturbed system
($a=0$) the sections show invariant curves $\gamma=$const. For
$a\neq 0$ some typical examples are shown in Fig.\ref{PNC1}.

 For small
values of $a$ (Fig.\ref{PNC1}a), the invariant curves are
perturbed slightly and only close to the most significant
resonances their deformation becomes noticeable. Increasing
further the perturbation parameter $a$, the width of the
resonances increases and homoclinic chaos becomes more obvious
close to the hyperbolic fixed points (Fig.\ref{PNC1}b). The
existence of invariant curves, which confine the resonant regions,
guarantees the bounded variation of the particle's energy
($\Delta\gamma=O(\sqrt a))$ for the chaotic trajectories.

When the amplitude $a$ exceeds a critical value $a_c$, overlapping
of resonances  takes place and large chaotic regions are generated
(Fig.\ref{PNC1}c) (see also \citet{Chirikov}). Particles with
initial energy $\gamma$ greater than a critical value $\gamma_c$
may follow a chaotic orbit which diffuse to regions of higher
energy, and this will lead them to very high energies in short
time scales. For relatively large values of $\alpha_c<<\alpha<1$,
the islands of regular motion, which survive from the resonance
overlapping, are gradually destroyed and chaos extends down to
relatively low energy particles (Fig.\ref{PNC1}d). The chaotic
part of the phase space will be  called ``the chaotic sea''.

\begin{figure}
\centering
\includegraphics[width=14cm]{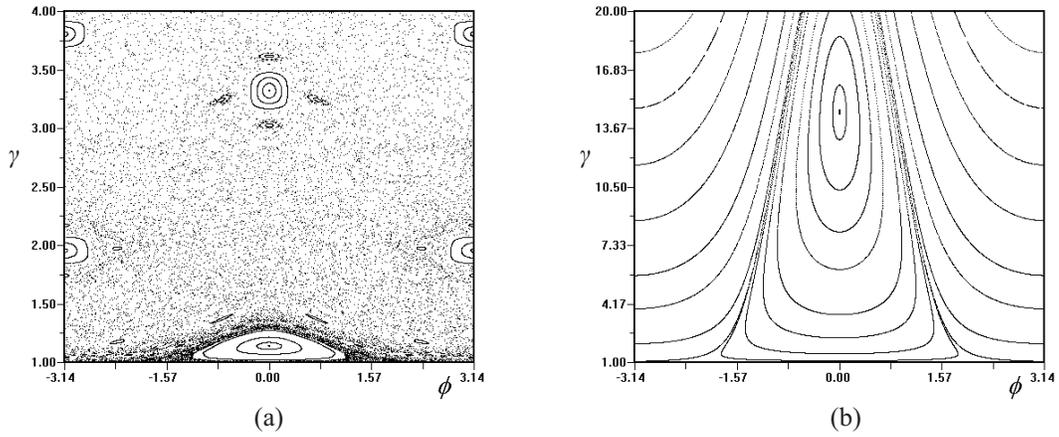}
\caption{Poincar\'e sections on the plane ($\phi,\gamma$) a)
$a=0.2, \nu=2, \theta=45^o$ b)$a=0.2, \nu=1, \theta=5^o$.}
\label{PNC2}
\end{figure}

The dynamics, presented by the Poincare sections in
Fig.\ref{PNC1}, is typical for the majority of parameter values.
Generally, the critical values $a_c$ and $\gamma_c$  determine the
conditions for possible chaotic diffusion. The dynamics of the
charged particles shows some exceptional characteristics when the
frequency of the GW is comparable to the Larmor frequency of the
unperturbed motion, particularly when $1\leq \nu < 3$. For such
parameter values, stochastic behavior  will appear when $\gamma=1$
and for sufficiently large perturbation values large chaotic
regions are generated and diffusion, even for particles with very
low initial energies, will be possible. An example is shown in
Fig.\ref{PNC2}a.

The evolution of the particles changes character when the
direction of propagation of the GW is almost parallel to $\vec B$.
In this case, chaos disappears, and the particles undergo  large
energy oscillations. As it is shown in Fig. \ref{PNC2}b, a
particle, starting even from rest ($\gamma\approx 1$), will be
driven regularly to high energies ($\gamma > 20$) and returns back
to its initial energy in an almost periodic way. In a realistic,
non infinite system, several particles may escape from the
interaction with the GW before returning back to low energies. At
$\theta=0$ the system is integrable and the energy of the
particles shows regular slow oscillations with an amplitude
proportional to $a$.

\section{Chaotic diffusion and particle acceleration}

\begin{figure}
\centering
\includegraphics[width=14cm]{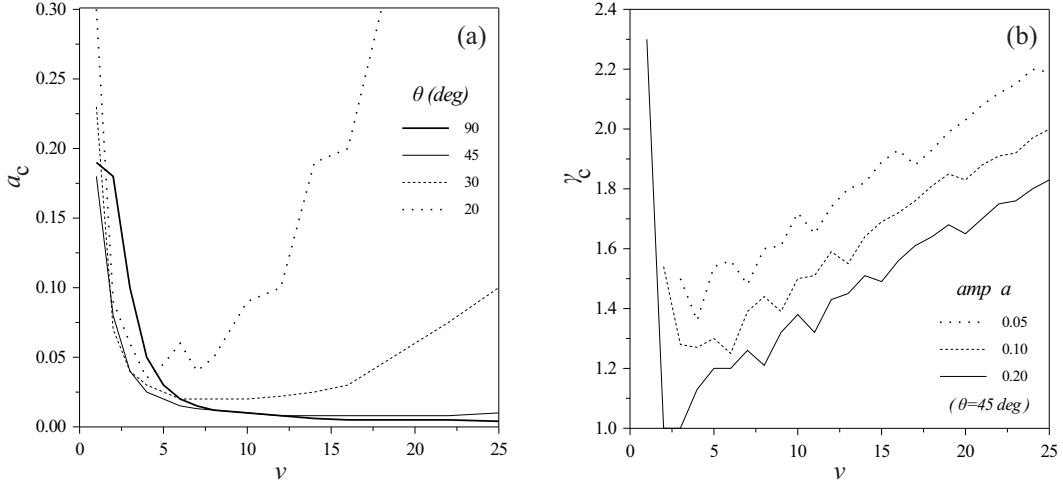}
\caption{a) The variation of critical value $a_c$ with the angular
frequency $\nu$ of the GW. $\;$ b) The same for the critical value
$\gamma_c$ and $\theta=45,$ using typical values  for the
amplitude $a$.} \label{ACGC}
\end{figure}

In the previous section, we showed that chaotic diffusion is
possible for $a\geq a_c$ and for the particles with $\gamma \geq
\gamma_c$. Such conditions are necessary but not sufficient for
acceleration,  since islands of regular motion may be present
inside the wide chaotic region (see for example Fig. \ref{PNC2}).
We estimate approximately the critical value $a_c$ by studying a
large number of Poincar\'e sections for different values of
$\theta$ (Fig.\ref{ACGC}a). We observe that for small values of
$\nu$ resonance overlap is obtained for relatively large
amplitudes of the GW. For higher frequency,  $\nu>5,$ and
$\theta>45$ the overlapping of resonances takes place even for
relatively small values of $\alpha$.  The critical value
$\gamma_c$ along the frequency axis $\nu$ is presented for
$\theta=45^o$ and for $a=0.05, 0.1, 0.2$ in Fig.\ref{ACGC}b. For
large values of $\nu$, and when the critical value $a_c$ is
relatively small ($\sim 0.005$), the critical particle energy
$\gamma_c$ is large and increases linearly with $\nu$. Namely, for
large $\nu$, chaotic diffusion takes place only for high energy
particles. On the other hand, for $\nu\approx 2$, resonance
overlapping takes place for large perturbations but the wide
chaotic sea formed extends down to the thermal velocity.

\begin{figure}
\centering
\includegraphics[width=14cm]{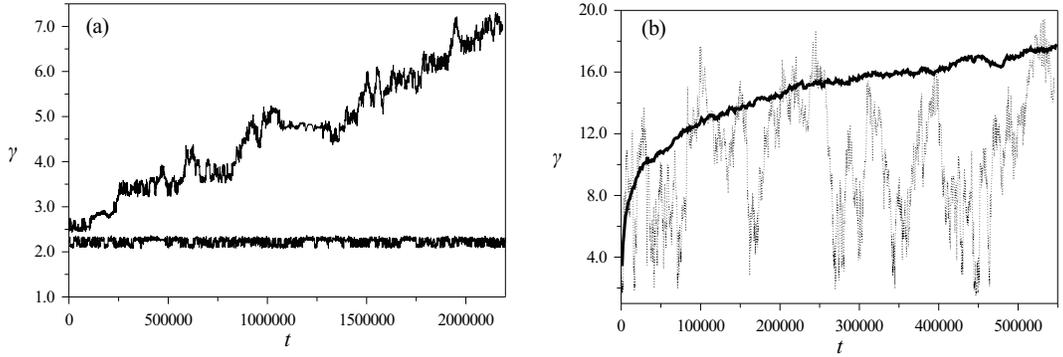}
\caption{a) The evolution of $\gamma$ along a trapped in a
magnetic island chaotic orbit for $\gamma(0)=2.2$,$\phi(0)=\pi$
and along a diffusive one for $\gamma(0)=2.6$, $\phi(0)=0$
($a=0.02, \theta=45^o, \nu=5$) b) The evolution of $\gamma$ along
a strongly chaotic orbit (dotted line) and its average value
(solid line) along 200 trajectories starting with $\gamma(0)=2.0$
and a randomly selected $\phi(0)$ ($a=0.1,\theta=45^o, \nu=5$).
The time is normalized with the gyro-period $(2\pi/\Omega)$.}
\label{GEVOL}
\end{figure}

In Fig. \ref{GEVOL}a the evolution of $\gamma$  along a
temporarily trapped chaotic orbit ($\gamma<\gamma_c$) and an orbit
which undergoes fast diffusion is shown, using $a=0.02$. In Fig.
\ref{GEVOL}b we plot the orbit of a particle which on the average
is not gaining energy and the  average rate of energy gain of 200
particles. The diffusion rate of the particles in the energy space
is initially fast but for time $t>5000$ it starts to slow down.
The time $t$ is normalized with the gyro period $2\pi/\Omega$

\subsection{GW interacting with an ensemble of electrons}
We study next the evolution of an energy distribution
$N(\gamma,t=0)$ of electrons interacting with the GW.

\begin{figure}
\centering
\includegraphics[width=16cm]{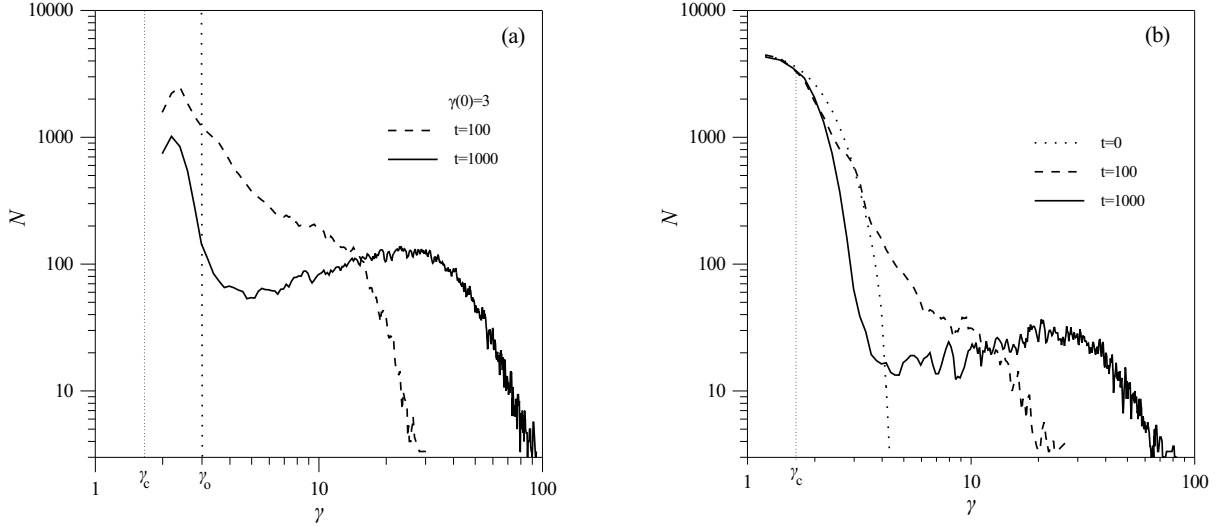}
\caption{The evolution of an energy distribution.  a) The initial
distribution (dotted line) consists of $3\times 10^4$ particles
having $\gamma(t=0)=3.$   b) The initial distributions is
Maxwellian, as it is shown by the dotted curve. Only particles in
the tail of the Maxwellian with $\gamma>\gamma_c$ will be
accelerated. The parameters used in both studies are  $a=0.5$,
$\nu=20$ and $\theta=30^o$.} \label{GDISTR}
\end{figure}

In Fig. \ref{GDISTR}a we follow the evolution of $3\times 10^4$
particles forming initially a cold energy distribution
$N(\gamma,t=0)\sim \delta(\gamma-3),$ where $\delta$ is the Dirac
delta function i.e.  all particles have the same initial energy
$\gamma=3.$  A large spread in their energy is achieved in short
time scales, and for $t=1000$, a non-thermal tail extending up to
$\gamma=100$ is formed.

We repeat the same analysis, assuming that the initial
distribution is the tail ($v>V_{the}$, where $V_{the}$ is the
ambient thermal velocity) of a Maxwellian distribution
(Fig.\ref{GDISTR}b).  The distribution of the high energy particle
form a long non-thermal tail analogously to the results reported
in Fig.(\ref{GDISTR}a).

\begin{figure}
\centering
\includegraphics[width=16cm]{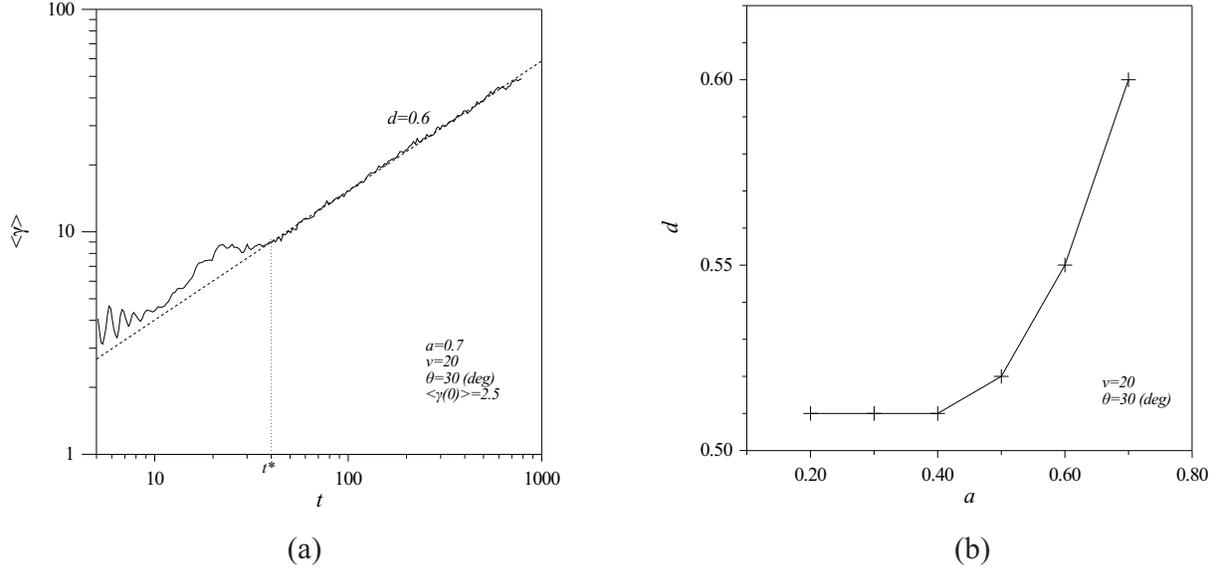}
\caption{(a) We plot the mean energy diffusion in time. We observe
that $\log (<\gamma>)$ is linearly related to $\log(t )$ for $t
> t^*$. The slope $d$ for $\alpha=0.7$ is $d\sim 0.6.$ (b) The slope
$d$ is plotted as a function of the GW amplitude.} \label{gamdif}
\end{figure}

The mean energy diffusion as a function of time is plotted in Fig.
\ref{gamdif}a for a particular set of parameters, and it has the
general form

\begin{equation}\label{diff}
    <\gamma> \sim t^d.
\end{equation}

From a large number of calculations, we find that the energy
spread in time follow a normal diffusion ($d=0.5$) in energy space
but as $\alpha$ increases (see Fig. \ref{gamdif}b), the
interaction becomes super-diffusive $(d \geq 0.5)$ in energy
space. This allows electrons to spread fast in energy space and
explains the efficient coupling between the GW and the plasma.

\section{A model for bursty acceleration}
\noindent Our main findings so far are:
\begin{enumerate}
    \item The GW can accelerate electrons from the tail of the ambient
    velocity distribution ($v_0(t=0)>V_{the}$) to very high energies
    ($\gamma >10-100$) for typical values of $0.005< \alpha < 0.5$
    and  $5\leq \omega/\Omega \leq 20.$
    Assuming that the the frequency of the
    GW is around 10KHz, and  $\Omega \sim
    \omega/3$ the magnetic fields strength  should be   around    $\sim 10^{-4}$ Gauss.
    \item The acceleration time depends in general on $\alpha$ but is relatively short
    $t_{acc} \sim 100 \Omega^{-1} \sim msecs$. During that time
    the GW will travel a distance $\ell_{acc} \sim c
    t_{acc} \sim 10^8$ cm.
    \item The mean energy diffusion of the electrons interacting with the GW
    follows the simple scaling $<\gamma>\sim t^d.$
    \end{enumerate}

\begin{figure}
\centering
\includegraphics[width=14cm]{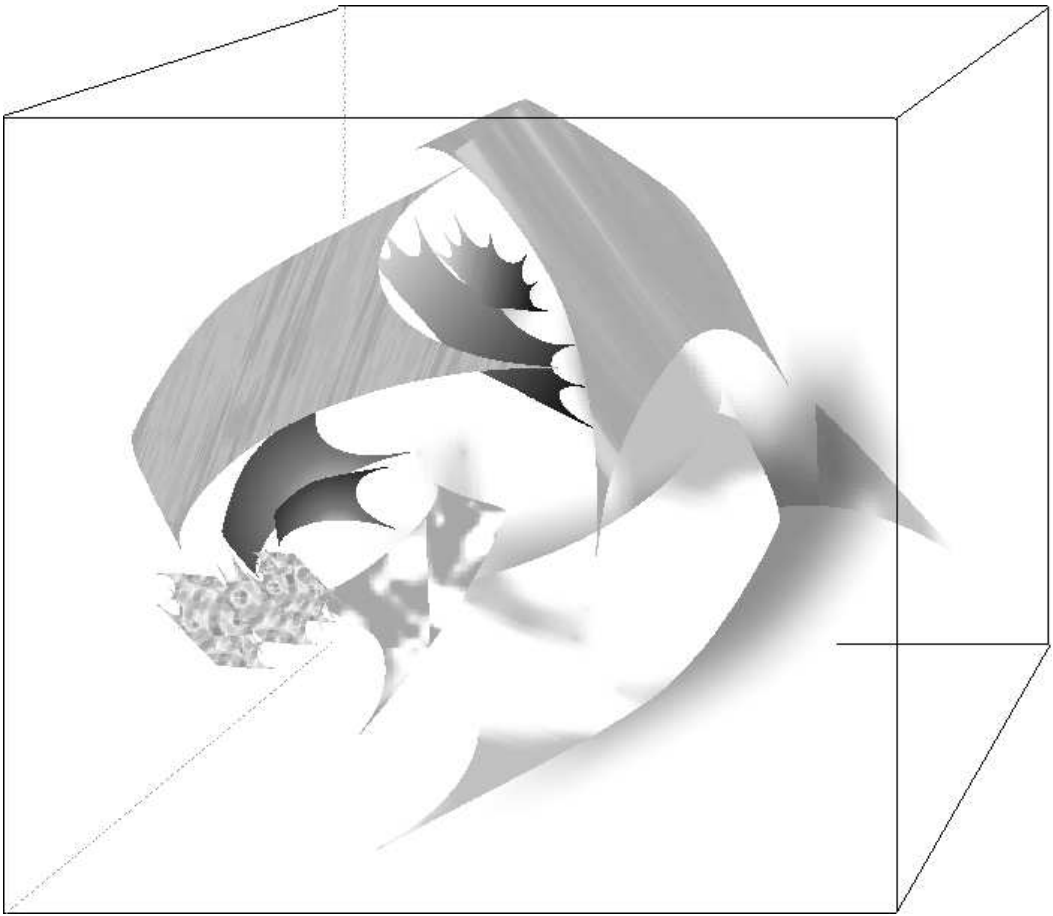}
\caption{A schematic representation of the thin three dimensional
magnetic null sheets appearing spontaneously and fill densely a
driven  turbulent magnetized plasma} \label{3dsheets}
\end{figure}

    We propose a new mechanism
    for efficient particle acceleration around strong and impulsive
    sources of GW using the estimates presented above
    for the strong interaction of GW
    with electrons. We assume that in the atmosphere of the central
    engine a turbulent magnetic field will be formed. Inside this
    complex    magnetic topology, a distribution of 3-D magnetic
    neutral sheets (magnetic null surfaces) with characteristic
    length $\ell$   will be
    developed (see Fig. (\ref{3dsheets}) and the relevant studies for the solar atmosphere
    \citep{Bul,Albi,Long,Long03}).    These 3-D surfaces naturally
    appear and disappear inside driven MHD turbulent
    plasmas and  are randomly distributed inside the atmosphere of the
    central engine. It is well known that
    the neutral sheets will act as localized dissipation
    regions and will accelerate particles \citep{Nodes}. The turbulent
    atmosphere of the central
    engine will sporadically emit weak X-ray and possibly $\gamma $-ray bursts
    for relatively
    long times  before the collapse    of the massive
magnetized star to form a black hole or the merging of neutron
star binaries. We suggest that numerous weak bursts are present
and remain unobserved.

\begin{figure}[ht]
\centering
\includegraphics[width=16cm]{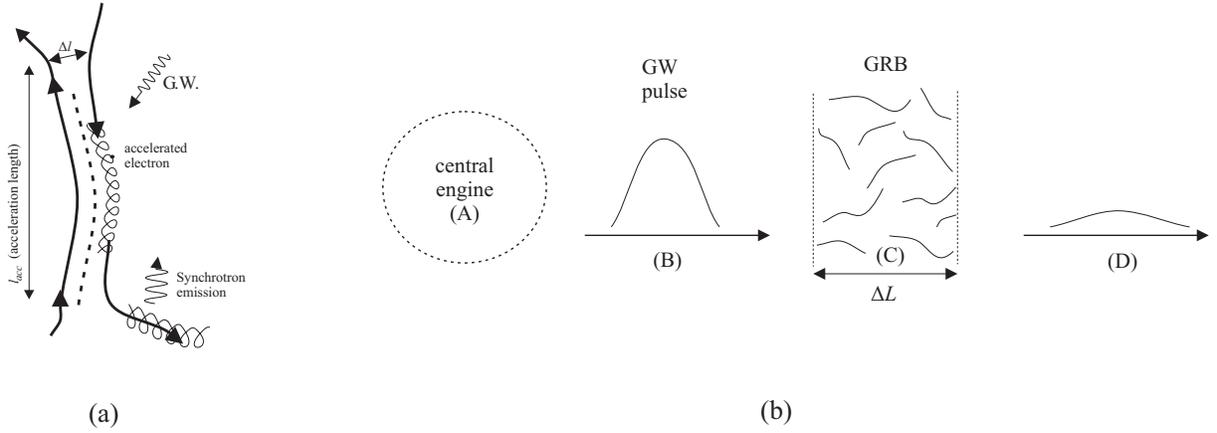}
\caption{(a) The propagation of a GW through a magnetic neutral
sheet accelerates electrons very efficiently. Relativistic
electrons stream away from the accelerator and emit a pulse of
synchrotron radiation when they reach the super strong magnetic
fields. The dashed line represents the magnetic field null surface
and $\ell_{acc}$ is the acceleration length.  (b) A collection of
magnetic neutral sheets is formed inside the turbulent atmosphere
(region C) of the central engine (region A). A GW pulse
propagating away from the central engine and passing through the
region C will form numerous $\gamma$-ray spikes by accelerating
particles near the magnetic null surfaces. The superposition of
these spikes form a short lived burst. The total burst duration is
approximately 100s ($\Delta T\sim  \Delta L/c$) but it is composed
by many short spikes lasting less than a second ($\ell/c$). The GW
pulse will become very weak and the density of the magnetic null
surfaces will drop dramatically in the region D, and this will
mark the end of the burst.} \label{MHDTUR}
\end{figure}

    In this article, we emphasize the role of the GW passing
    through magnetic neutral sheets and claim that   the GW
     will enhance dramatically the acceleration
     process inside the neutral sheet, causing very intense bursts. According to the
     arguments presented in sections 2 and 3, the GW can
     resonate with the electrons only when the magnetic field is
     weak (in the vicinity of the magnetic null surface). Inside
     the neutral sheet the synchrotron emission losses are
     negligible so the acceleration is very efficient.
     Electrons escaping from the neutral sheet to the super strong magnetic field,
     expected in the atmosphere of the
    central engine, transfer very efficiently their perpendicular
    to the magnetic field energy to synchrotron radiation creating spiky localized
    bursts. The relativistic electrons retain the
    parallel to the magnetic field energy
       and travel along the field lines till they
    reach the dense ambient plasma and deposit their energy
     via collisions,
      causing other
    types of longer lived bursts
    (e.g. X-ray and optical flashes)(see Fig. (\ref{MHDTUR}a)).

     We suggest thus that the creation of a network of localized
      accelerators with characteristic length $ \ell_{acc}$, which are spread in
    relatively large volumes inside the turbulent atmosphere
    of the central engine, will be responsible for the busty emission. The
    amplitude of the GW decays as it propagates away from the central engine
    as $\alpha \sim 1/r.$  When $\alpha <<a_c$ and/or the magnetic topology
    does not allow formation of magnetic neutral sheets the GW
    stops to interact with electrons and this indicates the end of the
    burst. So the overall duration of the burst is approximately
    $\Delta T\sim \Delta L/c$, where $\Delta L$ is the
    length of the characteristic layer where the interaction of
    the GW with the particles is efficient.
    (see Fig. \ref{MHDTUR}b).

\subsection{Energy spectrum}
    Let us now discuss another very important observational fact.
     It is well documented that the observed spectrum of the synchrotron
     radiation has a power law shape. This implies that
     the energy distribution causing the synchrotron
     radiation has also a power law  dependence in
     energy
     \[
     N(\gamma)\sim \gamma^{- \varepsilon}
     \]
     were $\varepsilon \simeq 1.8-2.0$ (e.g. see \citet{Sch} for GRB).
         These types of energy distribution are usually attributed to
     a shock wave (or a series of shock waves). The formation of
     a series of localized shock waves
     along the streaming plasma injected from the central engine
     is considered the ``standard'' model for GRB \citep{Piran, Mezaros}
     and may be of interest in many other astrophysical sources.

    The spontaneous formation of magnetic neutral sheets in the 3-D turbulent
     atmosphere of the central engine is not easy to describe.
    A number of studies on the statistical properties of the
    magnetic neutral sheets have appeared for the active region
    of the Sun \citep{geor,Long00,Isl00,Isl01,With02,Craig}.
    All these  studies are based on the observational fact
    that flares follow a specific
    distribution in energy \citep{cros}.

       An important concept from the study of
    3-D MHD turbulence is that a
    hierarchy of magnetic neutral sheets formed,  with a distribution of
    characteristic    lengths given by the function
 \begin{equation}\label{stat}
N(\ell) \sim \ell^{-b}.
\end{equation}
This scaling follows the energy distribution of flares and the
most probable value for $b$ is $3/2$. In other words, inside a
turbulent magnetized plasma, a large number of    magnetic neutral
surfaces with small characteristic lengths,
    and very few  large scale magnetic neutral
    sheets will be present.
    The exponent, according to the studies reported earlier,
     for the solar atmosphere is roughly equal to $b\sim
    1-2$ and is related to the statistics of X-ray bursts \citep{cros}.
The GW will accelerate electrons only in the brief period it
passes through the magnetic neutral sheet, so   the
    acceleration time will also follow a similar scaling law
    since
\begin{equation}\label{accel}
    t_{acc}\sim  \ell_{acc}/c.
\end{equation}
     We have shown in Fig. \ref{gamdif} that
    the the mean energy of the accelerated
    particles increases with time according to a simple power law
    as well.
Combining the Eq. (\ref{stat})-(\ref{accel}) we derive first the
distribution of $t$  which is a function of $\ell$. We have
$N(t)\,dt = N(l)\,dl$, or $N(t) = N(l) \frac{d\ell}{dt}$, which
yields, when neglecting the constants,
\begin{equation}
N(t) \sim t^{-b}. \label{Nt}
\end{equation}
We determine next the distribution of $\gamma$ (or, more
precisely, of $<\gamma>$), which is a function of $t$. We start
again from the relation $N(\gamma)\,d\gamma = N(t)\,dt$, or
$N(\gamma) = N(t) \frac{dt}{d\gamma}.$ We insert Eq.\ (\ref{Nt}),
$N(\gamma) = t^{-b} \frac{dt}{d\gamma},$ and we replace $t$ on the
right hand side using Eq.\ (\ref{diff}),
\[
N(\gamma) = (\gamma^{1/d})^{-b} \frac{d(\gamma^{1/d})}{d\gamma}
\]
or, on doing the derivative and rearranging,
\begin{equation}\label{enedist}
  N(\gamma) = \gamma^{(-b+1-d)/d}.
\end{equation}
Using a typical values for $d \sim 0.5$ (see Fig. \ref{gamdif})
and $b \sim 3/2,$ we estimate  the exponent of the energy
distribution to be around 2.

\subsection{Energetics}
The energy transferred from the GW to the high energy particles by
passing through one neutral sheet is

\begin{equation}\label{ener1}
  \mathbb{E}_{ns}\sim \left(\frac{n_t}{n_0}\right) n_0(\ell_{acc}^2\times \Delta \ell)\times
  \gamma mc^2
\end{equation}
where $n_0$ is the electrons density, $n_t$ are the electron at
the tail of the local Maxwellian, $\Delta \ell$ is the thickness
of the magnetic neutral sheet. The  filling factor of neutral
sheets inside the atmosphere of the central engine is assumed to
be $f.$  So the total number of magnetic neutral sheets expected
to interact with the GW is $\frac{(\Delta L)^3 f}{\ell^2\times
\Delta\ell}$ and the total energy transferred to the plasma
\begin{equation}\label{ener3}
  W_{tot}\sim \left[f\times \frac{(\Delta L)^3 }{\ell^2\times
\Delta\ell}\right ]\times
  \left(\frac{n_t}{n_0}
  n_0(\ell_{acc}^2\times \Delta \ell)\times
  \gamma mc^2\right).
\end{equation}
Using typical numbers $n_0=10^{12}\; \textrm{cm}^{-3}, n_t/n_0\sim
10^{-1}, \gamma=100, \ell_{acc}\sim 10^8\; \textrm{cm}, \Delta
\ell \sim 10^ 7 \;\textrm{cm} \;\; f \sim 10^{-1}$ and assuming
that the burst duration is 1s which implies that $\Delta L\sim
3\times 10^{12} \;\textrm{cm},$ the total energy estimated is
approximately $10^{47}-10^{48}$ ergs. The total burst  duration is
around 100s but the burst is composed with many short bursts
lasting less than a second ($\ell/c$).

    We can now list several characteristics of the bursty
    emission driven by the
    model    proposed above:

\begin{itemize}
\item A  fraction of the energy carried by the orbital energy of
the neutron stars at merger will go to the the GW and a portion of
this energy will be
    transferred    to the high energy electrons.
    \item The topology of the
    magnetic field varies from event to event, so every burst has
    its own characteristics.
    \item The superposition of many small scale localized sources
    produces a fine time structure on the burst.

    \item The superposition of  null surfaces with a power law
    distribution of the acceleration lengths will
    result in a power law energy distribution for the accelerated
    electrons and
    an associated synchrotron radiation emitted  by the relativistic electrons.
    \item The decay of the amplitude of the GW and/or
    the lack of magnetic neutral sheets away from
    the central engine will mark the end of the burst, but not
    necessarily the end of other types of bursts since
    the cooling of the ambient turbulent plasma has a much longer time scale.
\end{itemize}

It is worth drawing the analogy between the elements used to build
our model for this busrty emission and two significant
developments in solar flares and the Earths magnetic tail. (1)
\citet{Gar} studied the dynamic formation of current sheets
(current sheets are not always associated with neutral sheets) in
the solar atmosphere. They found that the irregular motion of the
magnetic foot-points, caused by the turbulent convection zone,
forces the magnetic topologies to form a hierarchy of 3-D current
sheets inside the complex magnetic topologies above active
regions. (2) \citet{Amb} studied the passage of low frequency
waves (Alfv\'en waves) through a 2-D neutral sheet. They
discovered that the acceleration of charged particles is greatly
enhanced from the presence of the Alfv\'en waves.

We believe that solar flares, the earths magnetotail, and the GW
driven bursts share two important characteristics, namely the
formation of 3-D magnetic null surfaces coupled with the passage
of waves through these surfaces (Alfv\'en waves for the sun and
the earths magnetotail and GW for the bursts analyzed in this
article). Although the external drivers
 and the
energetics of the bursts produced are clearly very different.

\section{Summary}
In this article we have study an efficient mechanism to transfer
the energy released from a central engine to the  turbulent plasma
in its atmosphere. We have analyzed the non-linear interaction of
GW with electrons. We showed that electrons in the tail of the
ambient plasma are accelerated efficiently, reaching energies up
to hundreds of MeV in less than a second, when (1) the amplitude
of the GW is above a typical value $0.005-0.5$ and (2) the
magnetic field is relatively weak $B_0\sim 10^{-4}$ Gauss, for a
GW with typical frequency 1-10KHz.

On the basis of these findings, we propose that pulsed GW emitted
from the central engine will interact with the ambient plasma in
the vicinity of the magnetic neutral sheets formed naturally
inside externally driven turbulent MHD plasmas. Magnetic neutral
sheets have characteristic lengths $\ell \sim 10^7-10^8$ cm and
are short lived 3-D surfaces. Although these structures are
efficient accelerators, we are emphasizing in this article only
the role of the GW passing through these surfaces since we focus
our attention on the very strong and bursty sources. The GW
passing through the neutral sheets will accelerate electrons to
very high energies. Relativistic electrons escape from the
magnetic neutral sheets radiating synchrotron emission as soon as
they reach the very strong magnetic fields.

 The collective emission of thousands of short lived
 (less than a second) synchrotron pulses,
created during the passage of the GW through a relatively large
volume. The relativistic electrons loosing most of their
perpendicular to the magnetic field energy to synchrotron
radiation retain the parallel energy and heat the ambient plasma
emitting other types of longer lived bursts, e.g. X-rays and
optical flushes.

We have also shown that if the acceleration length follows a
simple scaling law ($N(\ell_{acc}) \sim \ell_{acc}^{-b}$) and
since, as shown, $<\gamma >\sim t_{acc}^d,$ the energy
distribution also follows a power law scaling ($N(<\gamma >)\sim
<\gamma>^{(-b+1-d)/d}$), which for reasonable values of $b\sim
3/2$ and $d\sim 1/2$ agrees remarkably well with the energy
distributions inferred from the observations.

We are suggesting that numerous burst are still present before and
after the passage of the GW since the magnetic neutral sheet act
as efficient particle accelerators. The GW passage simply enhances
the acceleration process and makes many short lived spikes to
become visible. We propose that a more thorough analysis of the
statistical properties of the bursty emission, with main emphasis
on the low energy part of the frequency distribution, will give
important insight on the processes mentioned here.

A detailed model for the interaction of GW with turbulent MHD
plasma is currently under study, and we hope to develop  an even
more efficient energy transfer from the GW to the plasma e.g by
triggering the interaction (percolation) of many null sheets
during the passage of the GW. We hope that this may lead us to an
alternative scenario for the still unresolved questions related
with the acceleration mechanism in the atmosphere of the central
engines and the physical processes behind the X-ray and GRB.

\section*{Acknowledgments}

We thank Dr. Heinz Isliker and the anonymous referees for many
helpful suggestions which improved substantially our article.

\end{document}